%% file: proceedings.tex
\documentclass{PoS}

\usepackage{amsfonts,amsmath}

\title{Spatial diquark correlations in a hadron}

\ShortTitle{Spatial diquark correlations in a hadron}

\author{\speaker{Jeremy Green} and John Negele\\
        Center for Theoretical Physics, 
        Massachusetts Institute of Technology,
        Cambridge, MA 02139\\
        E-mail: \email{jrgreen@mit.edu}}

\author{Michael Engelhardt \\
  Department of Physics,
  New Mexico State University,
  Las Cruces, NM 88003-0001}

\author{Patrick Varilly \\
  Department of Physics,
  University of California, Berkeley,
  Berkeley, CA 94720}

\abstract{Using lattice QCD, a diquark can be studied in a
  gauge-invariant manner by binding it to a static quark in a
  heavy-light-light hadron. We compute the simultaneous two-quark
  density of a diquark, including corrections for periodic boundary
  conditions. We define a correlation function to isolate the
  intrinsic correlations of the diquark and reduce the effects caused
  by the presence of the static quark. Away from the immediate
  vicinity of the static quark, the diquark has a consistent shape,
  with much stronger correlations seen in the good (scalar) diquark
  than in the bad (vector) diquark.  We present results for
  $m_\pi=293$~MeV in $N_f=2+1$ QCD as well as $m_\pi=940$~MeV in
  quenched QCD, and discuss the dependence of the spatial size on the
  pion mass.  }

\FullConference{The XXVIII International Symposium on Lattice Field Theory\\
		 June 14-19,2010\\
		 Villasimius, Sardinia Italy}

\begin{document}

\section{Introduction}

Diquarks are two-quark systems. Collective behavior of a diquark has
been invoked to explain many phenomena of strong interactions
\cite{Anselmino:1992vg}. By introducing diquarks as effective degrees
of freedom in chiral perturbation theory, they have been used to
explain the enhancement of $\Delta I=\frac{1}{2}$ nonleptonic weak
decays \cite{Neubert:1991zd}. A simple quark-diquark model is quite
successful at organizing the spectrum of excited light baryon states
\cite{Selem:2006nd}.

The simplest diquark operators are quark bilinears with spinor part
$q^TC\Gamma q$. The favored combinations are color antitriplet, even
parity \cite{Jaffe:2004ph}. These are divided into ``good'' and
``bad'' diquarks.  The good diquarks, $q^TC\gamma_5q$, have spin 0 and
are flavor antisymmetric due to fermion statistics. The bad diquarks,
$q^TC\gamma_iq$, have spin 1 and are flavor symmetric.

Both one-gluon exchange in a quark model
\cite{DeRujula:1975ge,DeGrand:1975cf} and instanton
\cite{Schafer:1996wv} models give a spin coupling energy proportional
to $\vec{S_i}\cdot\vec{S_j}$, which favors the good diquark over the
bad diquark. The strength of this coupling falls off with increasing
quark masses. For the instanton model, the effective interaction has a
flavor dependence that also favors the good diquark.

\section{Earlier studies in baryons}

Since diquarks are not color singlets, studying them within the
framework of lattice QCD typically requires that they be combined with
a third quark to form a color singlet. Diquark attractions result in
spatial correlations between the two quarks in the diquark, which can
be probed by computing a wavefunction or two-quark density.

\begin{floatingfigure}[r]
\centering
\includegraphics[height=3cm]{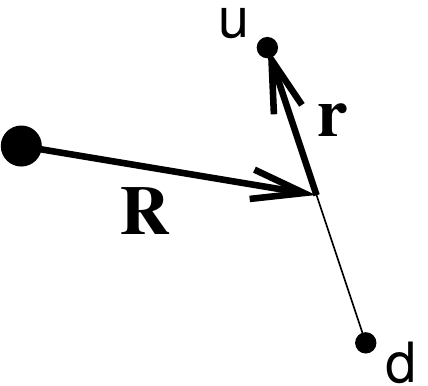}
\hspace{0.5cm}
\includegraphics[height=3cm]{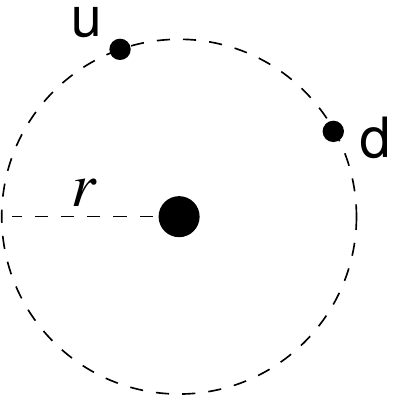}
\caption{\label{geometry}Geometry of the three quarks used in this paper and in \cite{Babich:2007ah} (left), and restricted geometry used in \cite{Alexandrou:2006cq} (right).}
\end{floatingfigure}

In one study \cite{Babich:2007ah}, using gauge fixing and three
quarks of equal mass, a wavefunction was computed by displacing
quarks at the sink,
\begin{align*}
 \psi(\mathbf{r}_1,\mathbf{r}_2) &\propto \sum_{\mathbf{r}_s} \langle
u(\mathbf{r}_s+\mathbf{r}_1,t)d(\mathbf{r}_s+\mathbf{r}_2,t)s(\mathbf{r}_s,t)\\[-0.8em]
&\qquad\qquad\times \bar
u(\mathbf{0},0)\bar d(\mathbf{0},0)\bar s(\mathbf{0},0)\rangle.
\end{align*}
Then, using the more convenient coordinates 
\begin{equation} \label{Rr}
\mathbf{R}=(\mathbf{r}_1+\mathbf{r}_2)/2 \quad \text{and} \quad
\mathbf{r}=(\mathbf{r}_1-\mathbf{r}_2)/2
\end{equation}
(Fig.~\ref{geometry}, left), the wavefunction of the good and bad
diquarks was shown for different fixed $R=|\mathbf{R}|$ as a function
of $\mathbf{r}$, in both Coulomb gauge and Landau gauge. In all cases,
the wave function had a peak near $\mathbf{r}=0$, but it was found to
fall off more rapidly for the good diquark, consistent with the
expectation that good diquarks are more tightly bound.

In a second study \cite{Alexandrou:2006cq}, spatial correlations were
investigated by computing the two quark density
$\rho_2(\mathbf{r}_u,\mathbf{r}_d)$ for a $(u,d)$ diquark in the
background of a static quark. To isolate correlations caused by the
diquark interaction, analysis was restricted to spherical shells
$|\mathbf{r}_u| = |\mathbf{r}_d| = r$ (Fig.~\ref{geometry},
right). For both good and bad diquarks, the density was found to be
concentrated near $r_{ud}=|\mathbf{r}_u-\mathbf{r}_d|=0$, and the
effect was much stronger for the good diquark. Fitting $\rho_2$ for
the good diquark to $\exp(-r_{ud}/r_0(r))$, $r_0$ reached a plateau
for large $r$, giving a characteristic size $r_0=1.1\pm0.2$~fm.

\section{Correlation function}

In the first study, the wavefunctions of good and bad diquarks were
compared for an unrestricted geometry, but they were not compared
against an uncorrelated wavefunction.
In the second study, the intrinsic clustering caused
by the diquark interaction was shown, however this was achieved by
using a restricted geometry.

To overcome these limitations, we combined a diquark with a static
quark, using the baryon operator $B = \epsilon^{abc}\left(u^T_aC\Gamma
  d_b\right)s_c$, taking $\Gamma=\gamma_5$ for the good diquark and
$\Gamma=\gamma_1$ for the bad diquark, and calculated the single quark
density and the simultaneous two-quark density:
\[
\rho_1(\mathbf{r}) = N_1\frac{\langle0|B(\mathbf{0},t_f)J_0^u(\mathbf{r},t)\bar{B}(\mathbf{0},t_i)|0\rangle}{\langle0|B(\mathbf{0},t_f)\bar{B}(\mathbf{0},t_i)|0\rangle}, \quad
\rho_2(\mathbf{r}_1,\mathbf{r}_2) = N_2\frac{\langle0|B(\mathbf{0},t_f)J_0^u(\mathbf{r}_1,t)J_0^d(\mathbf{r}_2,t)\bar{B}(\mathbf{0},t_i)|0\rangle}{\langle0|B(\mathbf{0},t_f)\bar{B}(\mathbf{0},t_i)|0\rangle}.
\]
Here, there are insertions of the current $J_\mu^f = \bar{f}\gamma_\mu
f$, and the normalization factors $N_{1,2}$ are required since this
local current is not conserved on the lattice.

In a system where $\rho_1(\mathbf{r})$ is uniform, the two-particle
correlation can be defined as
\[ C_0(\mathbf{r}_1,\mathbf{r}_2) = \rho_2(\mathbf{r}_1,\mathbf{r}_2)
- \rho_1(\mathbf{r}_1)\rho_1(\mathbf{r}_2). \] Deviations from zero
are seen as evidence for interactions between particles. This
correlation integrates to zero and approaches zero as the relative
distance $r_{12}=|\mathbf{r}_1-\mathbf{r}_2|$ increases beyond the
range of interactions in the system.

The situation considered here is not so simple. The single particle
density is not uniform: it is concentrated near the static
quark. $C_0$ will still integrate to zero and fall off at large
distances, however it is also larger near the static quark and this
obscures the diquark correlations.

In order to remove the effect of the static quark, we define the
normalized correlation function:
\begin{equation} \label{C}
C(\mathbf{r}_1,\mathbf{r}_2) = \frac{\rho_2(\mathbf{r}_1,\mathbf{r}_2)
                             - \rho_1(\mathbf{r}_1)\rho_1(\mathbf{r}_2)}
                            {\rho_1(\mathbf{r}_1)\rho_1(\mathbf{r}_2)}.
\end{equation}
This divides out the tendency to stay near the static quark and
retains the property of being zero if the two light quarks are
uncorrelated (i.e.\ if
$\rho_2(\mathbf{r}_1,\mathbf{r}_2)=\rho_1(\mathbf{r}_1)\rho_1(\mathbf{r}_2)$). The
downsides are that $C$ no longer integrates to zero, and it is
possible for $C(\mathbf{r},\mathbf{r})$ to increase without bound as
$|\mathbf{r}|\rightarrow\infty$.

\section{Density in a periodic box}

We assume the lattice spacing is small enough that in an infinite
volume we can treat $\rho_2(\mathbf{r}_1,\mathbf{r}_2)$ as a function
of $R=|\mathbf{R}|$, $r=|\mathbf{r}|$, and $\theta$, the angle between
$\mathbf{R}$ and $\mathbf{r}$. Since the calculation is actually
carried out on a finite lattice volume, to recover the infinite volume
result, we need to deal with the effect of periodic boundary
conditions.

The problem of dealing with $\rho$ in periodic boundary conditions has
been previously analyzed for the case of a meson
\cite{Burkardt:1994pw}. It was found that $\rho_1(\mathbf{r}) =
\sum_{\mathbf{n}\in\mathbb{Z}^3} \tilde\rho_1(\mathbf{r}+\mathbf{n}L),$ where
$\tilde\rho_1$ differs from the infinite volume result only for
$r\gtrsim L$ due to interactions with periodic images.

For this study, we have a baryon, and there is an additional
complication: the contribution from ``exchange diagrams'' in which the
two quarks travel in opposite directions across the periodic boundary
and can form a color singlet. As the lattice size grows, this becomes
dominated by the propagation of the lightest meson and so falls off as
$\exp(-m_\pi L)$.

Ignoring the exchange diagrams and interactions with periodic images
we find
\[
\rho_2(\mathbf{r}_1,\mathbf{r}_2)=\sum_{\mathbf{n}_1,\mathbf{n}_2\in\mathbb{Z}^3}\rho'_2(\mathbf{r}_1+\mathbf{n}_1L,\mathbf{r}_2+\mathbf{n}_2L),
\]
where $\rho'_2$ is the infinite volume two quark density.  In order to
deal with image effects, a phenomenological fit is used. Given a good
functional form $f'_2(\mathbf{r}_1,\mathbf{r}_2)$ for $\rho'_2$
(invariant under simultaneous rotations of $\mathbf{r}_1$ and
$\mathbf{r}_2$ as well as exchange of $\mathbf{r}_1$ and
$\mathbf{r}_2$), the nearest images are added~in:
\[
f_2(\mathbf{r}_1,\mathbf{r}_2)=\sum_{n_1^i,n_2^j\in\{-1,0,1\}}f'_2(\mathbf{r}_1+\mathbf{n}_1L,\mathbf{r}_2+\mathbf{n}_2L).
\]
This function and its lattice integral $f_1(\mathbf{r})=\sum_{\mathbf{r}_2}
f_2(\mathbf{r},\mathbf{r}_2)$ can be simultaneously fit to $\rho_2$ and
$\rho_1$ using a nonlinear weighted least squares method. This allows
the images to be subtracted off, giving $\rho'_2\simeq\rho_2-f_2+f'_2$
and $\rho'_1\simeq\rho_1-f_1+f'_1$, where $f'_1(\mathbf{r})=\int
d^3\mathbf{r}_2 f'_2(\mathbf{r},\mathbf{r}_2)$.

The so-called $\Delta$ ansatz for the static potential for interacting
quarks \cite{Alexandrou:2002sn,Takahashi:2002bw} was used as
motivation for the functional form of $f'_2$. We ultimately found that
the following eleven parameter functional form gave a reasonably
good fit:
\begin{gather*}
 f'_2(\mathbf{r}_1,\mathbf{r}_2) =  Ag(r_1,B,a_1,0)g(r_2,B,a_1,0)g(r,C,a_2,b_2) e^{-D\left(r_1^{3/2}+r_2^{3/2}\right)+Er^{3/2}+FR^{3/2}+Ge^{-\alpha\sqrt{r_1^2+r_2^2}}},\\
 \text{with}\quad g(r,A,a,b) = \begin{cases} \exp(-Ar) & r > a \\ c_1 - b r - c_2 r^2 & r <a \end{cases}, 
\end{gather*}
where $c_{1,2}$ are given by the requirement that $g$ and
$\frac{\partial g}{\partial r}$ are continuous at $r=a$.

\section{Lattice Calculations}

We used a mixed action scheme \cite{Bratt:2010jn} with domain wall
valence quarks on an asqtad sea, with $m_\pi=293(1)$~MeV and
$a=0.1241(25)$~fm. This ensemble had 453 HYP smeared MILC gauge
configurations \cite{Bernard:2001av}, which have $N_f=2+1$ and volume
$20^3\times 64$. Propagators were computed every 8 lattice units in
the time direction, allowing for 8 measurements per gauge
configuration, with source and sink separated by 8 lattice
units. Wilson lines were computed using HYP smeared gauge links, and
measurements were averaged over seven positions for the static quark:
$\mathbf{x}=\mathbf{0}$ and the six nearest neighbors.

For comparison, we also used a heavy quark mass, with $m_\pi\approx
940$~MeV. Since the effects of dynamical sea quarks are negligible at
that mass, we performed a calculation with $\kappa=0.153$ Wilson
fermions on 200 configurations from the OSU\_Q60a ensemble
\cite{Kilcup:1996hp}, which are $16^3\times 32$ with quenched
$\beta=6.00$ Wilson action. From the static quark potential, this has
$a/r_0=0.186$ \cite{Necco:2001xg}. Using $r_0=0.47$~fm, the
lattice spacing is $a = 0.088$~fm. We used a source-sink
separation of 11 lattice units and averaged measurements over the two
central timeslices.

The functions $f_{1,2}$ were fit to a restricted set of the lattice
measurements $\rho_{1,2}$. Three conditions were imposed to reduce the
influence of the points most affected by images: $r<8a$ for $\rho_1$,
$r_1^2+r_2^2<100a^2$ for $\rho_2$, and in both cases
$r_{\text{image}}\geq 11a$, where $r_{\text{image}}$ is the distance
to the nearest periodic image of the static quark. Fits had
$\chi^2$ per degree of freedom ranging from $0.25$ to $1.85$.

\begin{figure}[tbp]
\centering
\includegraphics[width=0.49\textwidth]{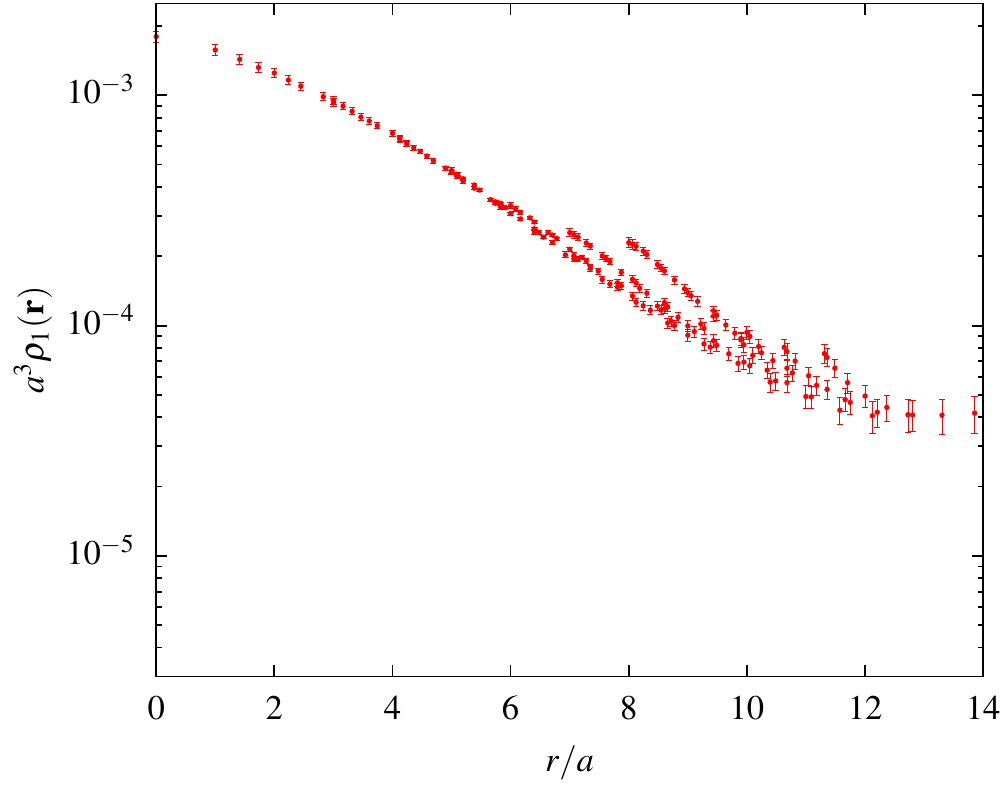}
\includegraphics[width=0.49\textwidth]{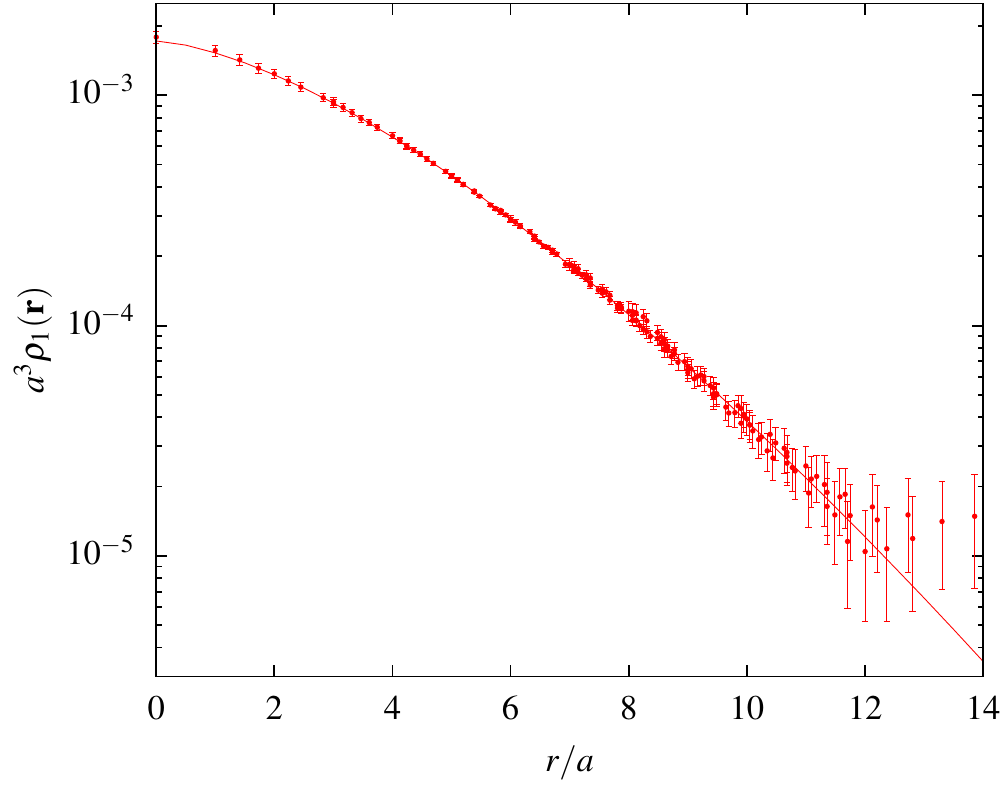}
\caption{\label{qgr1}$\rho_1(\mathbf{r})$ without (left) and with
  (right) image corrections for the good diquark on the quenched
  $m_\pi=940$~MeV ensemble}
\end{figure}

In the quenched good diquark case, Figure \ref{qgr1} shows the effect
of image corrections for $\rho_1$, and here this procedure is quite
successful, even extrapolating beyond the range included in the fit.

\begin{figure}[tbp]
\centering
\includegraphics[width=0.49\textwidth]{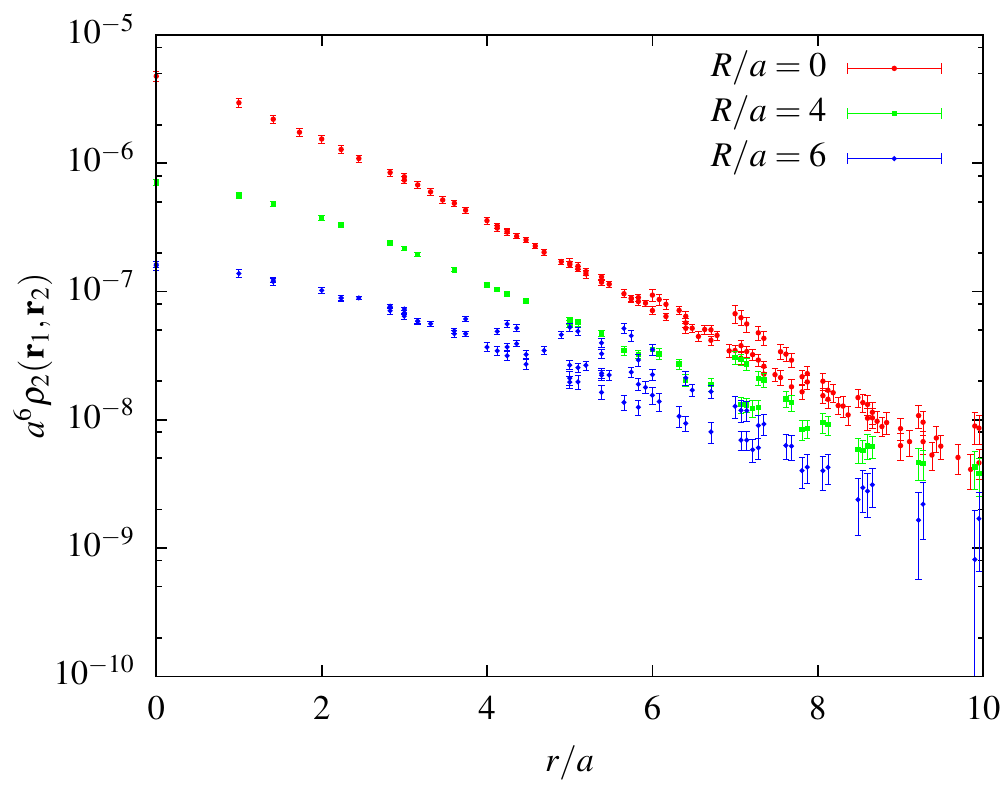}
\includegraphics[width=0.49\textwidth]{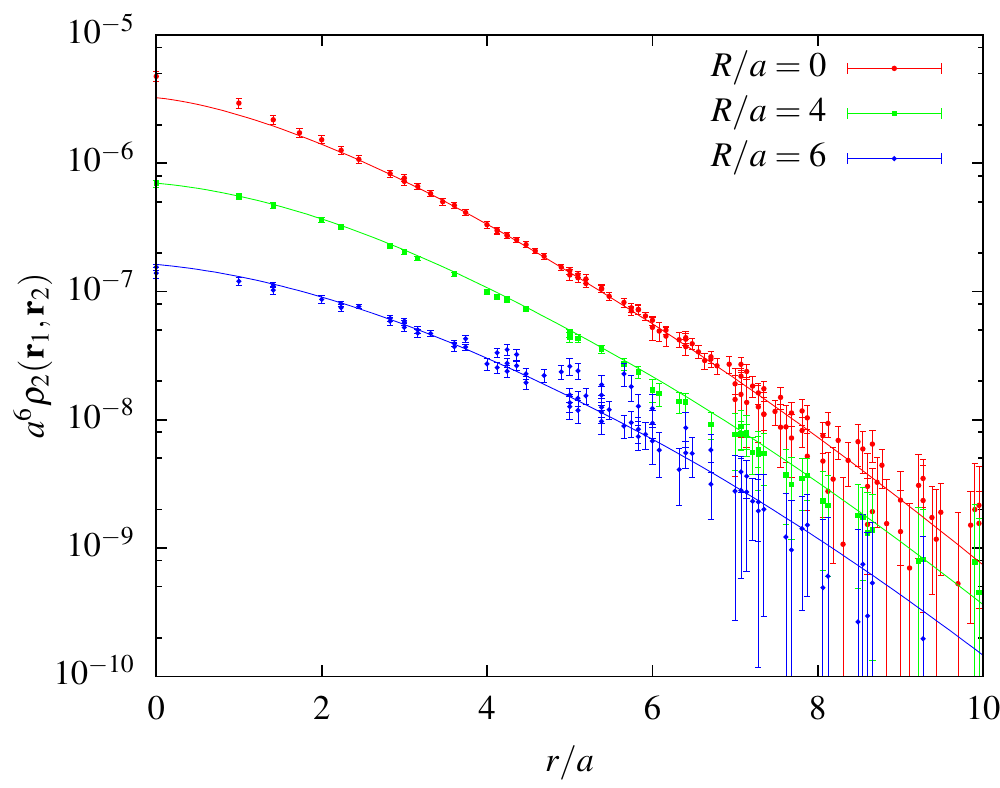}
\caption{\label{qgRt0}$\rho_2(\mathbf{r}_1,\mathbf{r}_2)$ without
  (left) and with (right) image corrections for the good diquark on
  the quenched $m_\pi=940$~MeV ensemble, as a function of $r$ with
  $\mathbf{r}\perp\mathbf{R}$ and $R/a=0,4,6$.}
\end{figure}

For $\rho_2$, the fit isn't as good as for $\rho_1$, but it still
works well. Figure \ref{qgRt0} shows $\rho_2$ with and without image
corrections.  The figure on the right looks cleaner for two
reasons. First, the fit function is determined using a global fit,
which allows for small deviation from the specified $R$ and $\theta$
to be compensated for. Second, image corrections have been applied,
which are substantial for points distant from the origin. The end
result is that the difference between the plotted point and the fit
curve is equal to the difference between the raw data point and the
fit for that point.

\section{Results and discussion}

As a check of how well the correlation function isolates the diquark
from the effect of the static quark, we compared different directions
of $\mathbf{r}$. Even at $R=0.2$~fm, $C$ was independent of the
direction of $\mathbf{r}$, indicating that this correlation function
works quite well.

\begin{figure}[tbp]
\centering
\includegraphics[width=0.49\textwidth]{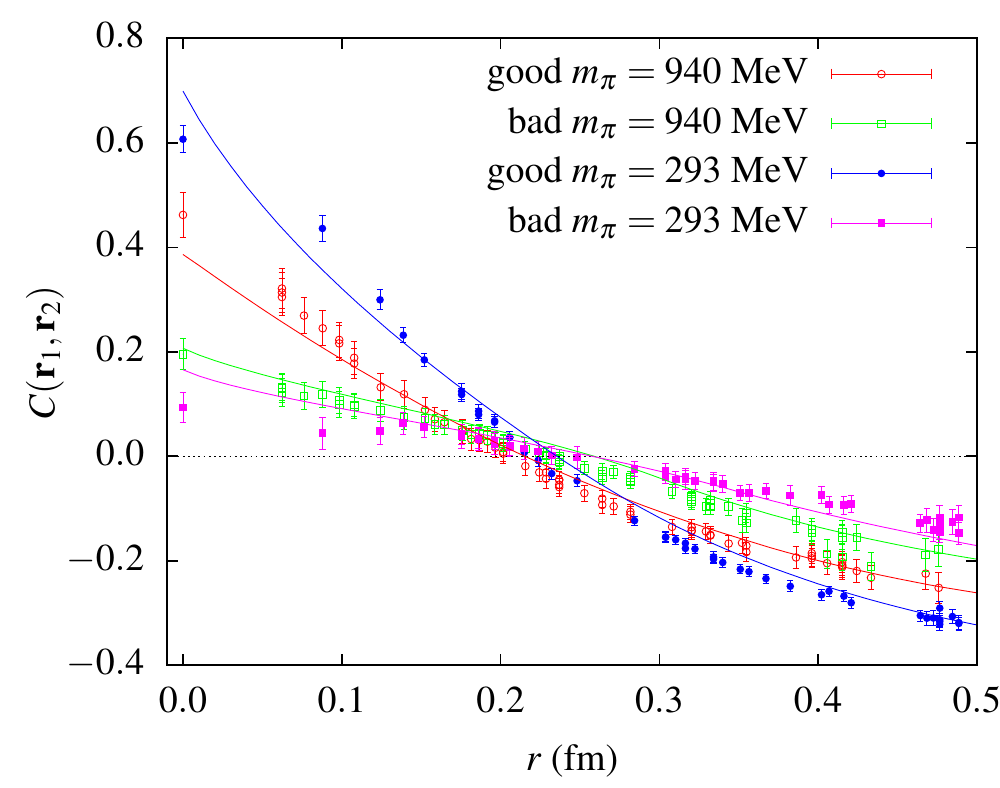}
\includegraphics[width=0.49\textwidth]{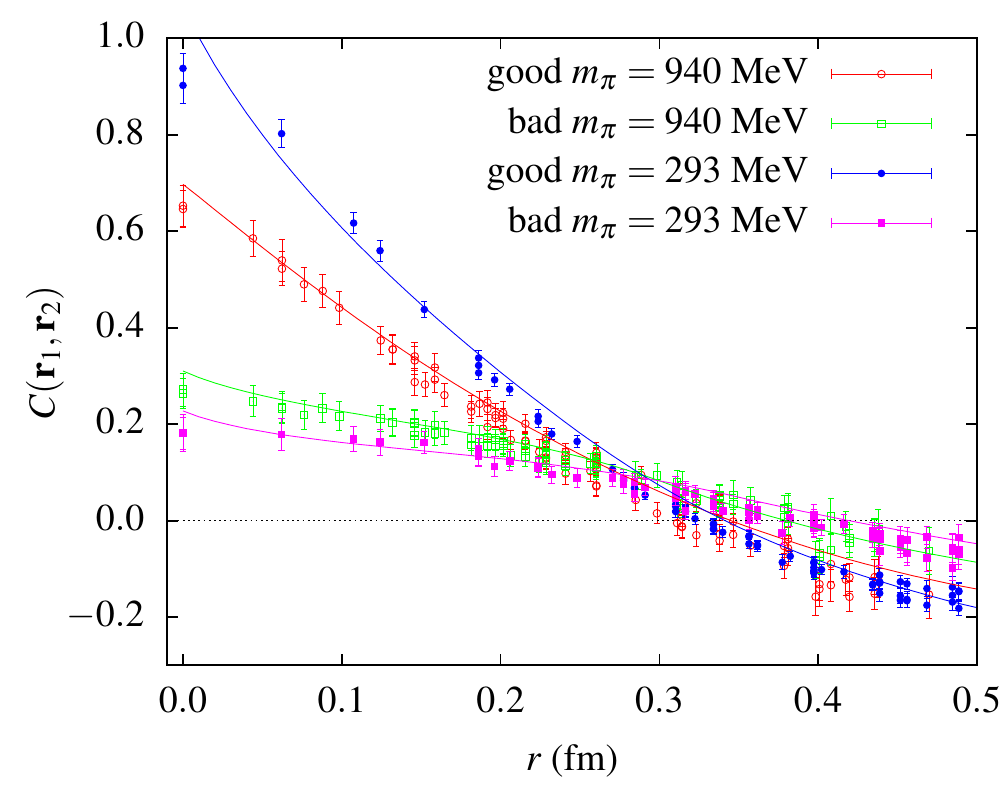}
\caption{\label{mCR0xt0r}$C(\mathbf{r}_1,\mathbf{r}_2)$, as a function
  of $r$~(in fm) with $R=0.2$~fm (left) and $R=0.4$~fm (right), with
  $\mathbf{r}\perp\mathbf{R}$ for the good and bad diquarks and the
  two pion masses.}
\end{figure}

\begin{figure}[tbp]
\centering
  \includegraphics[width=0.8\textwidth]{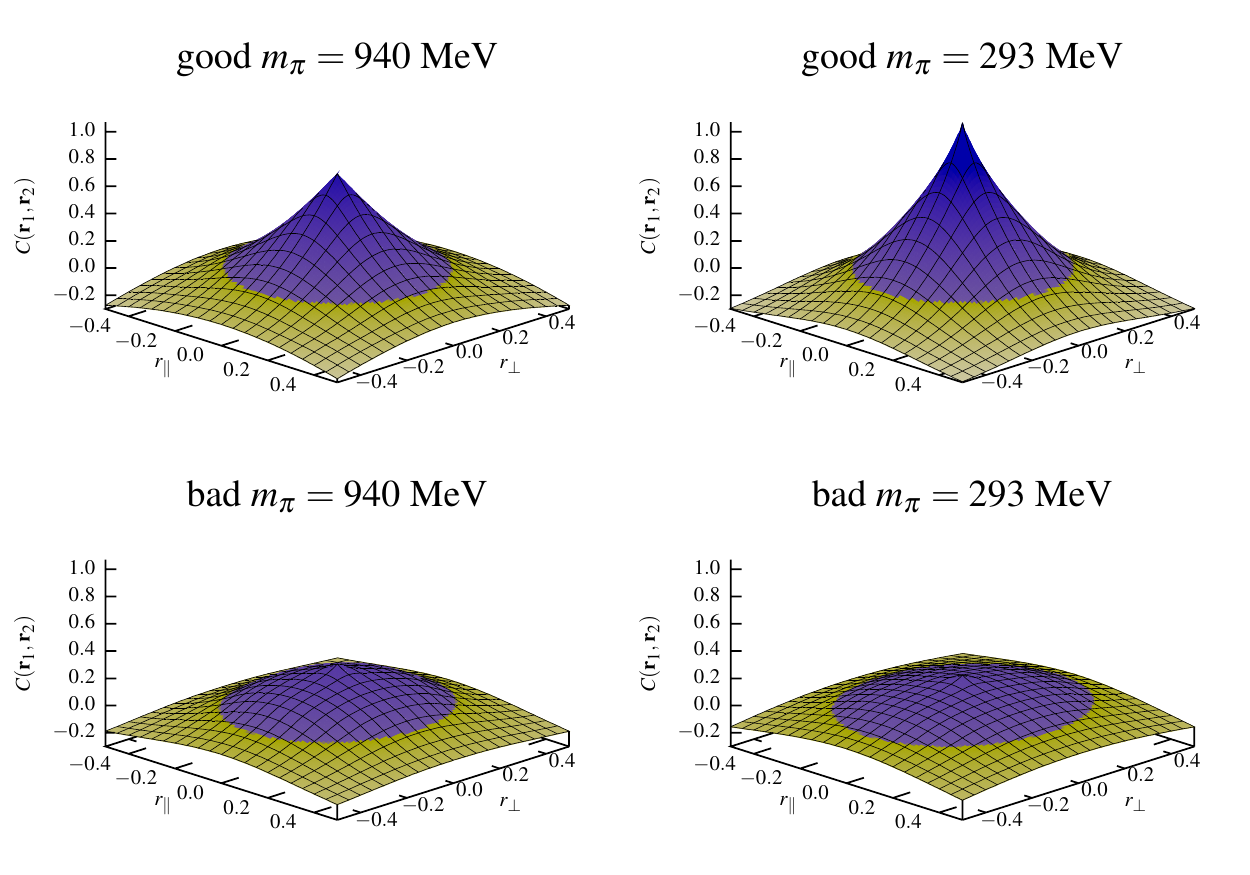}
  \caption{\label{plot3d}Continuous $C(\mathbf{r}_1,\mathbf{r}_2)$
    derived from the fit, as a function of $\mathbf{r}$~(in fm) with
    $R=0.4$~fm. The two axes $r_\parallel$ and $r_\perp$ indicate
    directions of $\mathbf{r}$ parallel to and orthogonal to
    $\mathbf{R}$, respectively. The color of the surface is
    discontinuous at $C=0$.}
\end{figure}

Finally, we can compare the systems.  Fig.~\ref{mCR0xt0r} shows the
profile of the correlation function $C$ at two fixed distances $R$
from the static quark to the center of the diquark, and
Fig.~\ref{plot3d} shows the full dependence of $C$ on $\mathbf{r}$, at
fixed $R=0.4$~fm.  The good diquark has a large positive correlation
at small $r$ that becomes negative at large $r$. The bad diquark has
similar behavior with smaller magnitude. The difference between the
good and bad diquarks is larger for the lighter pion mass, as expected
from the quark mass dependence of the spin coupling that splits good
and bad diquarks.  As $R$ increases, both the correlation and the size
of the positive region grow, although it is possible that some of this
growth of $C$ as $R$ increases may arise from the normalization of the
correlation function.

Our main conclusions are seen clearly in Fig.~\ref{plot3d}. The
diquark correlations are highly independent of $\theta$, indicating
negligible polarization by the heavy quark, are much stronger in the
good rather than the bad channel, and increase strongly with
decreasing quark mass. Finally, it is important to note that the
diquark radius is approximately $0.3$~fm and the hadron half-density
radius is also roughly $0.3$~fm, so the diquark size is comparable to
the hadron size. This is reminiscent of the size of Cooper pairs in
nuclei, and argues against hadron models requiring point-like
diquarks.

\acknowledgments

This work was supported in part by funds provided by the U.S.\
Department of Energy under Grants No.\ DE-FG02-94ER40818 and
DE-FG02-96ER40965. P.V.\ acknowledges support by the MIT Undergraduate
Research Opportunities Program (UROP). Additional domain wall
propagators were computed using the Chroma software suite
\cite{Edwards:2004sx}.

\input{proceedings.bbl}

\end{document}

%% file: proceedings.bbl
\providecommand{\href}[2]{#2}\begingroup\raggedright\endgroup